\documentstyle[epsfig,prc,aps,12pt]{revtex}
\begin{document}
%\draft
\tighten
\title{Breakup of $^{9}$Be on $^{209}$Bi above and near 
the Coulomb barrier as a
molecular single-particle effect: its influence on complete fusion and
scattering} 
\author{A. Diaz-Torres$^{a,}$$\footnote{Corresponding author.\\ 
{\it E-mail address}: A.Diaz-Torres@surrey.ac.uk\\
{\it Fax number}: +44 (0)1483 686781}$, 
I.J. Thompson$^{a}$ and W. Scheid$^{b}$}

\address{$^{a}$Physics Department, University of Surrey, Guildford GU2
7XH, United Kingdom\\
$^{b}$Institut f\"ur Theoretische Physik der Justus--Liebig--Universit\"at, 
D--35392 Giessen, Germany}

\date{\today}
\maketitle

\begin{abstract}

The breakup of the $^{9}$Be projectile on the $^{209}$Bi target at
bombarding energies above and near the Coulomb barrier is studied in the
adiabatic two-center shell model approach. The effect of $^{9}$Be$%
\rightarrow $n$+2\alpha $ breakup channel on complete fusion, elastic and
inelastic cross sections is investigated. Results show that the breakup of
the projectile $^{9}$Be could be due to a molecular single-particle effect
shortly before the colliding nuclei reach the Coulomb barrier.
\end{abstract}


\section{Introduction}

\subsection{Review of the subject}

\smallskip The interaction of weakly bound or halo nuclei with stable
targets at colliding energies around the Coulomb barrier is a very lively
topic due to the increasing interest in nuclear reactions with radioactive
beams. The existence and the role of the breakup process of weakly bound
projectiles in complete fusion and scattering (elastic and inelastic)
mechanisms have been extensively investigated in recent years both
theoretically \cite{TH2,TH4,TH1,Yabana,Dasso,hagino} and experimentally \cite
{Yoshida,Takahashi,Rehm,Signorini1,Kolata,Dasgupta,Trotta,Aguilera,Signorini2,Lubian,Mukherjee}%
, but there is not yet any definitive conclusion.

There are contradictory theoretical works which predict either the
suppression \cite{TH2,TH4,TH1,Yabana} or the enhancement \cite{Dasso} of the
complete fusion cross section due to the coupling of the relative motion of
the colliding nuclei to the breakup channel. In a recent paper \cite{hagino}%
, coupled channels calculations for $^{11}$Be+$^{208}$Pb have shown that the
coupling of the relative motion to the breakup channel has two effects
depending on the value of the bombarding energy, namely the reduction of the
complete fusion cross sections at energies above the Coulomb barrier due to
the loss of incident flux and the enhancement of the complete fusion cross
sections at energies below the Coulomb barrier due to the dynamical
renormalisation of the nucleus-nucleus potential. In \cite{TH2,TH4,TH1}, the
effects of the breakup process on fusion were studied in the framework of
both time-independent and time-dependent semi-classical coupled channels
theories using the imaginary part of a local dynamical polarisation
potential associated with the breakup channel. In \cite{Yabana}, the
influence of the breakup process on fusion was studied by solving a
three-body Schr\"{o}dinger equation with a time-dependent wave packet
method. The breakup process and its effects on fusion were described in \cite
{Dasso,hagino} in terms of inelastic excitations of the projectile to the
continuum within the coupled channels formalism.

In experiments of $^{9,10,11}$Be+$^{209}$Bi \cite{Yoshida,Signorini1} and $%
^{9}$Be+$^{208}$Pb \cite{Dasgupta}, it was observed that the breakup process
of both $^{9}$Be and $^{11}$Be projectiles suppresses the complete fusion
above the Coulomb barrier. Reactions of $^{6,7}$Li + $^{9}$Be also reveal
the same hindrance for fusion \cite{Takahashi}, although contrary results 
also exist, e.g., $^{9}$Be+$^{9}$Be \cite{Mukherjee}. This fusion
hindrance was not observed in the reaction $^{9}$Be+$^{64}$Zn \cite{Lubian}.
We expect that the presence or absence of a suppression mechanism in fusion
reactions involving $^{9}$Be could be associated with the evolution of
the last neutron of $^{9}$Be in the approach phase of the collision before 
the nuclei reach the Coulomb barrier, which
depends on the structure of the target or projectile. We also expect that
molecular effects, which will be discussed later, do not occur in some
reactions with $^{9}$Be. 
From an optical model analysis of the $^{9}$Be+$^{209}$Bi elastic scattering
\cite{Signorini2}, the imaginary (absorptive) potential at the distance
corresponding to the strong interaction radius increases with decreasing
bombarding energy towards the Coulomb barrier energy, but this is not the
case for the system $^{9}$Be+$^{64}$Zn \cite{Lubian}. This behaviour of the
imaginary potential indicates that strong absorption channels are still open
for bombarding energies very close to the Coulomb barrier and it was
associated with the Coulomb breakup of $^{9}$Be. Moreover, it was pointed
out that other absorption channels related to orientation and deformation
effects of $^{9}$Be cannot be excluded. Such effects have been reported for
the scattering of $^{9}$Be by $^{40,44}$Ca and $^{39}$K \cite{Hnizdo}.

\subsection{Object of the work}

The aim of this paper is to study the breakup of the $^{9}$Be projectile on
the $^{209}$Bi target at energies above and near the Coulomb barrier in the
adiabatic two-center shell model (TCSM) \cite{TCSM} approach, as well as its
role in the complete fusion and scattering processes. The TCSM is based on 
the assumption that the nucleons can be described with molecular states 
during the heavy ion reaction. This means that the motion of the nuclear 
centers has to be adiabatically slow compared with the rearrangement of 
the mean field of nucleons. Whether molecular orbitals are formed depends 
on the ratio of two characteristic times, 
the collision time $\tau_c \sim 2R/v= 2R(2E/MA)^{-1/2}$ and 
the nuclear period or single-particle 
relaxation time $\tau_s \sim 2R/v_{Fermi}=2R(2\epsilon_F/M)^{-1/2}$, where 
$E$ is the laboratory bombarding energy, $A$ the projectile mass number, and 
$\epsilon_F$ the Fermi energy in the target or projectile nucleus 
with valence nucleons. According to the ratio of $\tau_c/\tau_s \sim$ 3-4 
for the studied reaction, we expect that molecular orbitals and 
the polarisation of valence nucleons by the field of the other nucleus may 
have time to develop.   

We study the neutron
level diagram in the approach phase of this collision in order to see
possible single-particle effects (nucleon transfer or single-particle
excitations). Since the $^{9}$Be has a small separation energy ($S_{n}=1.67$
MeV), the breakup channel ($^{9}$Be$\rightarrow $n+$^{8}$Be) could be
related to a diabatic single-particle motion of the last neutron of $^{9}$Be
into the continuum \cite{Cassing} or to a neutron transfer to $^{209}$Bi.
Since the nucleus $^{8}$Be has no bound states, as the Coulomb field of the
target ($^{209}$Bi or $^{210}$Bi) in its vicinity is very strong and the
relative motion is considered adiabatic, 
we assume that the decay $^{8}$Be$\rightarrow
2\alpha $ occurs automatically and immediately. 
This assumption is experimentally supported by the observation of $\alpha $ 
particles originating from the decay of $^{8}$Be \cite{Bologna}. 
The capture of a part of the
projectile after the breakup is associated with the incomplete fusion
channel which will not be considered in the present work for the sake of
simplicity.

In general, the neutron promotion from $^{9}$Be to $^{209}$Bi or into the
continuum could be related to three types of transitions \cite{FM1} between
the molecular adiabatic single-particle levels: the transition induced by
the radial relative motion of nuclei (radial transition $\sim d/dr$), the
transition induced by the rotation of the internuclear axis around a fixed
axis in the laboratory system (rotational transition $\sim jI$ , $j$ and $I$
being the single-particle total angular momentum and the total angular
momentum of the system, respectively), or the transition induced by the
change of the orientation of the intrinsic symmetry axes of deformed nuclei,
which is dependent on the angular momenta of the nuclei ($\sim I_{1}^{\prime
}$, $I_{2}^{\prime }$ with $I_{1}^{\prime }+I_{2}^{\prime }=I-L$ and $L$
being the orbital angular momentum of the collision). In the TCSM \cite{TCSM}
used in the present work, only radial and rotational transitions between the
molecular adiabatic single-particle levels can occur because the intrinsic
symmetry axis of nuclear shapes lies parallel to the internuclear axis. For
arbitrary orientations of the intrinsic symmetry axes of deformed nuclei, a
more general TCSM \cite{Nuhn} would have to be used. The radial transition
occurs between adiabatic states with the same angular momentum projection $%
j_{z}$ and is largest at the point of an avoided crossing (pseudo-crossing),
while the rotational transition occurs between adiabatic states $\phi
_{a},\phi _{b\text{ }}$with $j_{z}^{a}=j_{z}^{b}\pm 1$ and is largest at the
crossing point of levels. Moreover, the radial excitation mechanism
describes the non-adiabaticity of the relative motion (Landau-Zener
mechanism \cite{LZ}) and grows with increasing radial velocity, the diabatic
single-particle motion being a limit for large radial transition
probabilities \cite{Cassing}. In the diabatic limit of the single-particle
motion \cite{Cassing}, the nucleons do not occupy the lowest free
single-particle levels as in the adiabatic case, but remain in the diabatic
levels keeping their quantum numbers during a collective motion of the
nuclear system.

The study of single-particle level diagrams \cite{FM1} can be useful to
understand some reaction channels and to simplify coupled channels
calculations. A microscopic formulation of the excitation and nucleon
transfer in the TCSM and its role in the scattering process of nuclei with
loosely bound nucleons was presented in \cite{Park} on the basis of the
molecular particle-core model. A similar formulation was suggested in \cite
{Imanishi} in the framework of the orthogonalised coupled-reaction-channel
theory. Molecular single-particle effects are experimentally well
established phenomena \cite{FM1} in heavy ion collisions and have been
reflected, e.g., in large structures observed in elastic and inelastic
excitation functions of reactions between light nuclei.

In sect. 2, trajectories, complete fusion, elastic and inelastic cross
sections are calculated in the framework of a classical model, using the
potential energy obtained with the adiabatic TCSM along with
phenomenological friction coefficients. The role of the breakup of $^{9}$Be
on complete fusion, elastic and inelastic cross sections is incorporated by
using the Landau-Zener approach for the breakup induced by the radial
mechanism. We compare the calculated results with experimental data in sect.
3. A summary and conclusions are given in sect. 4.

\section{Model}

\subsection{Adiabatic TCSM}

The basic microscopical model for a dinuclear system is the TCSM. The TCSM
describes nuclei, namely the united nucleus and the two incoming nuclei. In
the TCSM \cite{TCSM} used in the present work, the nuclear shapes are
defined by a set of coordinates:

\begin{enumerate}
\item  The relative distance $r$ between the centers of mass of the two
nuclei (or the elongation $\lambda $ of the system, or the distance $%
(z_{1}-z_{2})$ between the centers of the two oscillators).

\item  The neck parameter $\varepsilon =E_{0}/E^{\prime }$ is defined by the
ratio of the actual barrier height $E_{0}$ to the barrier height $E^{\prime
} $ of the two-center oscillator.The neck grows with decreasing $\varepsilon 
$.

\item  The transition of the nucleons through the neck is described by the
mass (charge) asymmetry $\eta =(A_{2}-A_{1})/A$ ($\eta _{Z}=(Z_{2}-Z_{1})/Z$%
), where $A_{1}(Z_{1})$ and $A_{2}(Z_{2})$ are the mass (charge) numbers on
both sides of the neck and $A=A_{1}+A_{2}$ ($Z=Z_{1}+Z_{2}$).

\item  The deformations $\beta _{i}=a_{i}/b_{i}$ ($i=1,2$) of axial
symmetric fragments are related to the ratio of their semiaxes. For
separated nuclei, the semiaxes $a_{i}$ and $b_{i}$ can be related to the
parameter of quadrupole deformation by using the known expansion of the
nuclear surface in spherical harmonics.
\end{enumerate}

In the adiabatic approach, the parameters of the momentum-independent part
of the single-particle TCSM potential are determined by the assumption of
conservation of the volume enclosed by the nuclear surface, continuity of
the potential and its derivatives between the two centres and by the
requirement that the TCSM potential barrier with height $E_{0}$ is located
at $z=0$. The parameters related to the momentum-dependent part of the
potential (spin-orbit and $l^{2\text{ }}$- like terms) are determined by
some interpolation between the appropriate values for the compound nucleus
and for the separated fragments for large $r$. In the present work, the same
interpolation method as in \cite{TCSM} is adopted, but the parameters $%
\kappa $ and $\mu $ of the Nilsson model for the spin-orbit interaction are
used \cite{Ring}. The nuclear-radius constant is r$_{0}=1.2249$ fm and the
oscillator quanta are $\hbar \omega _{i0}=41\cdot A_{i}^{-1/3}$ MeV.

The single-particle levels, characterised by the projection $j_{z}$ of the
total angular momentum, are given in the harmonic oscillator potentials only
up to constants $V_{i}$ associated with the depths of the two potential
wells. In order to compensate this disadvantage, we use three parameters
related to the depths of potential wells of the compound nucleus $V_{0}$ and
the separated nuclei $V_{i}^{\infty }$ for large $r$. These parameters are
expressed by the sum of experimental nucleon separation energies \cite
{Sn1,Sn2,Sn3} and corresponding Fermi levels. For finite relative distances,
we interpolate the values of the depth of potential wells $V_{i}$ as

\begin{equation}
V_{i}=\left\{ 
\begin{array}{c}
V_{0}+(V_{i}^{\infty }-V_{0})\frac{\omega _{z}^{i}-\omega _{z}^{0}}{\varpi
_{z}^{i}-\omega _{z}^{0}},\text{ if }R_{neck}\text{ }>0 \\ 
V_{i}^{\infty },\text{ if }R_{neck}\text{ }=0
\end{array}
\right\} ,  \label{well}
\end{equation}
where $\varpi _{z}^{i}$ are the frequencies of the oscillator along the
internuclear $z$ axis for the configuration of the system with a radius of
the neck $R_{neck}$ $=0$. The $\omega _{z}^{i}$ and $\omega _{z}^{0}$ are
the $z$-oscillator frequencies for the nuclei in the united system and for
the compound nucleus, respectively. It is necessary to point out that $V_{i}$
are added to the single-particle energies taking care to identify every
level in association with its localisation in only one of the two potential
wells for separated nuclei.

In order to avoid unrealistic shapes for small relative distances $r$
(compact shapes) and a large mass asymmetry $\eta $ of the colliding nuclei
in this version of the TCSM, it is assumed that the mass asymmetry $\eta $
decreases, e.g. linearly \cite{TCSM}, with decreasing relative distance $r$
from the touching configuration of the nuclei. An adequate shape
parametrisation for such highly asymmetric compact configurations was used
to study cluster emission in the superasymmetric TCSM \cite{Mirea}. Since
the motion of the system is mainly considered in the entrance channel up to
relative distances near the touching configuration, the parametrisation used
for compact shapes does not influence the results. Moreover, compact shapes
of the system do not occur in the approach phase of a collision for
bombarding energies above the Coulomb barrier owing to the repulsive core of
the nucleus-nucleus potential caused by the centrifugal effect and by
diabatic \cite{Diabatic2,Diabatic3} or structural forbiddenness effects \cite
{Adamian1}.

The orbital angular momentum independent part of the total potential of the
two-center system is calculated with the macroscopic-microscopic method of
Strutinsky as 
\begin{equation}
\!V=E_{macr}(r,\varepsilon ,\eta ,\eta _{Z},\beta _{1},\beta _{2})+\delta
E_{shell}(r,\varepsilon ,\eta ,\eta _{Z},\beta _{1},\beta _{2})+\delta
E_{pair}(r,\varepsilon ,\eta ,\eta _{Z},\beta _{1},\beta _{2})-V_{CN}.
\label{stk}
\end{equation}

The smooth macroscopic part of the potential energy $E_{macr}$ is expressed
by the sum of the Coulomb energy $E_{C}$ and the nuclear surface energy $%
E_{N}$ as in the liquid drop model, and they contain a part related to the
binding energy of the system and another one related to the nucleus-nucleus
interaction potential. In order to take into consideration the finite range
of the nuclear force and the diffuse nuclear surface, we use the nuclear
surface energy $E_{N}$ obtained by the Yukawa-plus-exponential method \cite
{Krappe} with shapes generated by the TCSM potential. The fluctuating
microscopic shell and pairing corrections, $\delta E_{shell}$ and $\delta
E_{pair}$, are obtained using the single-particle levels of the TCSM and the
Strutinsky method. The dependence of these microscopic corrections on
temperature is neglected because the dissipated energy for the studied
reaction is only a few MeV \cite{Signorini2}. The value of the potential (%
\ref{stk}) is normalised to the potential energy $V_{CN}$ of the spherical
compound nucleus.

\subsection{Classical trajectories}

Classical trajectories for a system with reduced mass $\mu ,$ total energy $%
E $ and orbital angular momentum $L=\mu r^{2}d\varphi /dt$ in the center of
mass system, are obtained by solving the generalised Lagrange equations in
the relative distance $r$ and orientation $\varphi $ ($q_{i}=\{r,\varphi \}$%
) of the two centers of the nuclei

\begin{equation}
\frac{d}{dt}\left[ \frac{\partial \pounds }{\partial (dq_{i}/dt)}\right] -%
\frac{\partial \pounds }{\partial q_{i}}=F_{i}(q,dq/dt)\equiv
-\sum_{j}\gamma _{ij}\stackrel{}{f_{j}(q)dq_{j}/dt},  \label{lagrange}
\end{equation}
where $\pounds =T-V$ is the Lagrangian and $F_{i}$ are the friction forces.
As in the surface friction model \cite{Gross1}, we assume a diagonal
friction tensor $\gamma _{ij}=c_{i}^{0}\delta _{ij}$ (radial and tangential
friction) with friction form factors $f_{j}(q)$. The form factor of the
friction forces is calculated as\smallskip 
\begin{equation}
f_{r,\varphi }=[\nabla (E_{N}-\widetilde{E}_{N})]^{2},  \label{formf}
\end{equation}
$\widetilde{E}_{N}$ being the nuclear surface energy for the configuration
of the system with a radius of the neck $R_{neck}=0$. We have assumed in (%
\ref{formf}) that frictional effects start with the overlapping of the two
nuclei. Theoretical justifications for the classical form of the friction
forces were given in \cite{Gross1}, where the perturbation theory was used
to study the coupling between the collective and the intrinsic degrees of
freedom for small overlap between the nuclei. The tangential part of the
friction is considerably weaker than the radial part \cite{Gross1}. The
phenomenological friction forces in the classical trajectory express the
irreversible coupling between the relative motion of nuclei and the degrees
of freedom that are not explicitly considered (e.g., collective vibrations
of $^{209}$Bi), for small overlap between the nuclei outside the touching
distance. Moreover, this phenomenological friction does not take into
account the reversible coupling (diabatic single-particle motion due to the
Landau-Zener mechanism) between the relative motion and the intrinsic
degrees of freedom \cite{Mosel1}, which is much stronger for large overlap
between the nuclei inside the touching distance. The friction parameters $%
c_{r}^{0}$ and $c_{\varphi }^{0}$ are usually considered as free parameters
and their values are determined from extensive analyses of fusion cross
sections and inelastic scattering data for many systems and bombarding
energies throughout the periodic table \cite{Gross1,Fobrich1}. For
simplicity, we assume that the reduced mass $\mu $ and the friction
coefficients $c_{i}^{0}$ do not depend on the temperature of the system.
Moreover, the slight increase of the reduced mass with decreasing relative
distances for weakly necking shapes around the touching configuration is
neglected \cite{Adamian2}. In \cite{Dietrich}, it was pointed out that the
inclusion of this contribution does not lead to any appreciable change of
the observables in the exit channel. For large overlap between the nuclei,
these transport coefficients can be calculated in an adiabatic approach
within the linear response theory \cite{Hofman2}. In the present work,
classical trajectories are calculated using the radial and the tangential
friction coefficients $c_{r}^{0}=4\times 10^{-23}$ MeV$^{-1}$s and $%
c_{\varphi }^{0}=0.01\times 10^{-23}$ MeV$^{-1}$s, respectively \cite{Gross1}%
.

\subsection{Complete fusion, elastic and inelastic cross sections}

The solution of the classical equations of motion (\ref{lagrange}) leads to
captured and scattered (elastic and inelastic) trajectories. Since the
initial mass asymmetry $\eta $ of the system ($^{9}$Be+$^{209}$Bi) is larger
than its Businaro-Gallone mass asymmetry \cite{Diabatic3}, we assume that
the complete fusion channel is mainly related to the relative distance $r$
between the centers of the nuclei. In general, the complete fusion cross
section can be expressed as

\begin{equation}
\sigma _{fus}=\sum_{l=0}^{l_{cr}}\sigma _{cap}(\text{E},l)\cdot P_{CN}(\text{%
E},l),  \label{fus1}
\end{equation}
where $\sigma _{cap}$, $P_{CN}$ and $l_{cr}$ are the partial capture cross
section, a hindrance factor for complete fusion and the highest orbital
angular momentum (a sharp angular momentum cut-off model) for trajectories
which reach the critical distance $r_{cr}$, respectively. At the critical
distance $r_{cr},$ the shells of the individual nuclei dissolve into those
of the compound nucleus \cite{Mosel1}. Expression (\ref{fus1}) is valid
if the mechanism hindering the complete fusion occurs before or after
the Coulomb barrier has been passed. The partial capture cross section $%
\sigma _{cap}$ is defined as 
\begin{equation}
\sigma _{cap}=\frac{\pi \hbar ^{2}}{2\mu \text{E}}(2l+1)\cdot T_{l}(\text{E}%
),  \label{fus2}
\end{equation}
where E is the bombarding energy. The transmission coefficient $T_{l}$
through the Coulomb barrier $V_{Bl}$ with orbital angular momentum $L=$ $%
l\hbar $ is approximated by that of a parabolic barrier with angular
momentum dependent frequency $\hbar \omega _{l\text{ }}$

\begin{equation}
T_{l}(\text{E})=\frac{1}{1+\exp [2\pi (V_{Bl}-\text{E})/\hbar \omega _{l}]}.
\label{fus3}
\end{equation}
%%\qquad \qquad \qquad \qquad \qquad

In (\ref{fus1}), the factor $P_{CN}$ takes into account other hindrances for
the compound nucleus formation. In the reaction $^{9}$Be$+^{209}$Bi, we
assume that the hindrance is mainly related to the breakup of the projectile 
$^{9}$Be (e.g., $^{9}$Be$\rightarrow $n+$2\alpha $):

\begin{equation}
P_{CN}=1-P_{bup}(\text{E},l),  \label{fus9}
\end{equation}
where $P_{bup}$ is the breakup probability. The hindrance factor (\ref{fus9}) 
was used in \cite{TH2,TH4,TH1} and suggested experimentally 
in \cite{Signorini1,Dasgupta}. 
The role of $P_{bup}$ in the complete fusion cross section depends on 
the existence of the breakup channel which
is derived, in the present model, from the diagram of neutron levels 
calculated with the TCSM.
Since for orbital angular momenta $0\leq l\leq l_{cr}$ the breakup
probability $P_{bup}$ changes much less than the partial capture cross
section $\sigma _{cap}$, we use an average value $\overline{P_{bup}}$ with
respect to orbital angular momenta. The complete fusion cross section (\ref
{fus1}) is factorised as

\begin{equation}
\sigma _{fus}=\left[ \frac{\pi \hbar ^{2}}{2\mu \text{E}}\sum_{l=0}^{l_{cr}}%
\frac{2l+1}{1+\exp [2\pi (V_{Bl}-\text{E})/\hbar \omega _{l}]}\right] \cdot
(1-\overline{P_{bup}}(\text{E})).  \label{fus11}
\end{equation}

For simplicity, we introduce the approximations \cite{Mosel} $\hbar \omega
_{l}=\hbar \omega $ and $r_{Bl}=r_{B}$, with $\hbar \omega $ and $r_{B}$
being the frequency and the position of the $l=0$ Coulomb barrier $V_{B}$,
respectively. Moreover, the summation in Eq. (\ref{fus11}) is replaced by an
integration and the term in brackets yields an analytical expression as the
formula for the complete fusion cross section in the Glas-Mosel model \cite
{Mosel}. The Glas-Mosel model is used, instead of a dynamical fusion model
as the dynamical energy surplus model \cite{Ngo}, to calculate complete fusion cross sections
without breakup because we found that the radial kinetic energy
loss of nuclei ($^{9}$Be+$^{209}$Bi) before reaching the Coulomb barrier
is negligible compared with the static Coulomb barrier $V_{B}$, e.g., 
$\leq 0.1$ MeV at
the bombarding energy $1.15V_{B}$ (E$_{\text{c.m.}}=46$ MeV) 
for $0\leq l\leq l_{cr}$. However, frictional effects on the radial motion 
of nuclei are important in calculating the breakup probability $P_{bup}$
because this probability exponentially depends
on the radial velocity of nuclei as we will discuss further on.

In the complete fusion process, the probability for a breakup process
induced by the Landau-Zener mechanism at an isolated pseudo-crossing of
adiabatic single-particle levels can be expressed as

\begin{equation}
P_{bup}=P_{LZ}^{(e)},  \label{LZ4}
\end{equation}
and in the scattering process as

\begin{equation}
P_{bup}=P_{LZ}^{(e)}+(1-P_{LZ}^{(e)})\cdot
P_{LZ}^{(lv)},  \label{LZ3}
\end{equation}
since the system passes through the pseudo-crossing point twice, first
entering $(e)$ and then leaving $(lv)$ the interaction region. Here, we have
made a distinction between $(e)$ and $(lv)$ because the radial velocity
changes due to frictional effects. 
The first term in Eq.(\ref{LZ3}) is $P_{LZ}^{(e)}$, instead of 
$P_{LZ}^{(e)}\cdot (1-P_{LZ}^{(lv)})$, because of our assumption of
an immediate breakup of $^{8}$Be $\rightarrow 2\alpha $ after the neutron in
$^{9}$Be has been removed from the nucleus entering the interaction region 
(the neutron does not get bound again with $^{8}$Be when the nuclei leave 
the interaction region). 
At an isolated pseudo-crossing of
adiabatic single-particle levels $E_{1},E_{2\text{ }}$for a relative
distance $r_{0}$, the Landau-Zener probability \cite{LZ} for a nucleon
transition between the single-particle levels is

\begin{equation}
P_{LZ}=e^{-2\pi G^{0}},  \label{LZ1}
\end{equation}
with 
\begin{equation}
G^{0}=\frac{\mid H_{12}^{\prime }\mid ^{2}}{\hbar v^{0}\mid \frac{d}{dr}%
(\epsilon _{1}-\epsilon _{2})\mid _{r=r_{0}}},  \label{LZ2}
\end{equation}
where $\mid H_{12}^{\prime }$ $\mid $ is a coupling matrix element between
the diabatic states $\epsilon _{1},\epsilon _{2}$ and is equal to half the
closest distance between the adiabatic levels $E_{1},E_{2\text{ }}$at $r_{0}$%
. Diabatic states $\epsilon _{1},\epsilon _{2}$ are a set of states which
run smoothly through the crossing. The radial velocity $v^{0}$, which is the
only quantity concerning relative motion in the above expression, can be
calculated from the classical equations of motion (\ref{lagrange}). Since
the radial velocity $v^{0}$ depends on the incident energy E and the orbital
angular momentum $l$, $P_{LZ}$ depends also on E and $l$. The Landau-Zener
formula is derived under the following assumptions: $H_{12}^{\prime }$ is
smaller than the relative kinetic energy of the nuclei, $(\epsilon
_{1}-\epsilon _{2})$ is a linear function of $r$, and $v^{0}$ is constant.
The restriction to constant velocities implies that the point of avoided
level crossing is far from the turning point of the relative motion of the
two nuclei. The assumption of a constant relative velocity can be
approximately fulfilled in a region where the nucleus-nucleus potential is
shallow. In general, the neutron transition at a pseudo-crossing can be
calculated semiclassically \cite{FM1} using a time-dependent Schr\"{o}dinger
equation for the neutron wave function expanded in a basis of two diabatic
wave functions related to the diabatic levels $\epsilon _{1},\epsilon _{2}$
and a classical equation for the radial relative motion of the nuclei.

The scattering function $\theta (l)$ is calculated from the classical
trajectories. Elastic and inelastic scattering cross sections $\sigma $ are
obtained from the classical scattering cross section $\sigma _{cl}$ as

\begin{equation}
\frac{d\sigma }{d\Omega }=\frac{b(\theta )}{\sin \theta }\left| \frac{%
db(\theta )}{d\theta }\right| \cdot P_{sc},  \label{scat2}
\end{equation}
where the classical scattering cross section $\sigma _{cl}$ is given in
terms of the scattering angle $\theta $ and of the impact parameter $%
b=l\hbar /\sqrt{2\mu \text{E}}$. The hindrance factor for scattering $%
P_{sc}=1-P_{bup}$ ($P_{bup}$ is obtained from Eq. (\ref{LZ3})) takes into
account the fact that the number of scattered projectiles is reduced by the
breakup process. We define as elastic events those non-captured trajectories
which show less than $0.5$ MeV energy loss, and as inelastic ones all other
non-captured trajectories.

\section{Results of the calculation and discussion}

\smallskip Since the collision above the Coulomb barrier is discussed, we
first consider $^{9}$Be and $^{209}$Bi as spherical nuclei (ratio of the
semiaxes $\beta _{i}=1$), and then study the deformation effects. The
assumption of spherical nuclei is suitable if the change of the
nucleus-nucleus potential, due to orientation and deformation effects of the
nuclei, is small in comparison with the surplus of bombarding energy above
the potential. It is known \cite{Fobrich1} that the deformation effects in
the entrance channel may be important for collisions very close to the
Coulomb barrier, where they mostly cause an increase of the complete fusion
cross section.

In order to obtain a nuclear shape for the touching configuration of the
nuclei similar to the one supplied by the overlap of the two nuclear frozen
densities \cite{Diabatic2,Diabatic3,Transp3}, the neck parameter $%
\varepsilon $ should be set at about $0.75$. With this value of $\varepsilon 
$, the neck radius and the distance between the centers of the nuclei are
approximately equal to the corresponding quantities in the dinuclear system
formed by the overlap of the two frozen densities.

\subsection{Neutron level diagram of the adiabatic TCSM}

Fig. 1 shows the neutron level diagram of the TCSM for the system $^{9}$Be + 
$^{209}$Bi $\rightarrow $ $^{218}$Fr as a function of the relative distance $%
r$ between the spherical nuclei. For $r\approx 5.5$ fm (the relative
distance between the centers of mass of the two hemispheres of the spherical
compound nucleus), we recognise the neutron single-particle states of the
spherical compound nucleus $^{218}$Fr and for large $r$ the neutron
single-particle states of the spherical separated nuclei $^{9}$Be and $%
^{209} $Bi. The quantum numbers $l$,$j$ for these asymptotic neutron levels
are shown. The single-particle levels quickly change in the vicinity of $%
r\approx 5.5$ fm because of the abrupt change of the nuclear shape from a
spherical shape (at $r\approx 5.5$ fm) to compact asymmetric shapes with a
large value of the neck parameter $\varepsilon =0.75$ (for $r\gtrsim 5.5$
fm). Small values of the neck parameter $\varepsilon $ ($0-0.2$) are more
realistic for such compact shapes near the spherical shape because the
potential energy of the system decreases with decreasing values of $%
\varepsilon $ \cite{FM1}. The nuclei $^{9}$Be and $^{209}$Bi only reach
spherical shapes for large $r$ due to a polarisation effect caused by the
smoothing of the barrier which joins the two oscillator potential wells \cite
{TCSM}. This evolution to spherical shapes depends slightly on the value of
the neck parameter $\varepsilon $. In Fig. 1, L$^{*}$ and H$^{*}$ denote the
highest occupied level for the nuclei $^{9}$Be and $^{209}$Bi, respectively.
The large gap for the neutron subsystem of $^{209}$Bi associated with the
neutron magic number N=126 is observed. The internal arrow (A) indicates the
distance $r_{t}=9.82$ fm corresponding to the touching configuration of the
nuclei and the external one (C) indicates the relative distance $r=13.15$ fm
for the configuration of the system with a radius of the neck $R_{neck}$ $=0$%
. The Coulomb barrier (B) is located between these two arrows at $r_{B}=11.4$
fm.

In Fig. 1 at $r\approx 12$ fm, we can see two very close pseudo-crossings
(denoted by 1 and 2) between the highest occupied state of $^{9}$Be with $%
j_{z}=3/2$ and two unoccupied states of $^{209}$Bi with $j_{z}=3/2$.
Pseudo-crossings (denoted by 3 and 4) between these states also occur at
relative distances close to the touching configuration of the nuclei. The
level denoted by L corresponds to a level of $^{9}$Be with $j_{z}=1/2$,
occupied by two neutrons. This level shows a pseudo-crossing (denoted by 5)
at $r\approx 12.5$ fm with an unoccupied state of $^{209}$Bi with $j_{z}=1/2$
as well as a crossing (denoted by 6) with another state of $^{209}$Bi with $%
j_{z}=3/2$. In general, at pseudo-crossings 1-5 transitions induced by a
radial mechanism can occur. The rotational mechanism can induce a transition
at crossing 6, although this type of transition is much weaker than the
radial one \cite{FM1}. The radial transitions 1-4 are associated with the
most favourable breakup channel $^{9}$Be $\rightarrow $n$+2\alpha $ ($%
Q=-1.57 $ MeV).

In the present work, we will only consider transitions at the
pseudo-crossings 1-2 of Fig. 1. The last neutron of $^{9}$Be could be
promoted into the continuum at pseudo-crossing 1 and could be transfered to $%
^{209}$Bi at pseudo-crossing 2. At the pseudo-crossings 1-2, the
applicability of the Landau-Zener approach is quite good. The values of $%
\mid H_{12}^{\prime }\mid $ and $\frac{d}{dr}(\epsilon _{1}-\epsilon _{2})$
at the pseudo-crossings 1-2 are $0.1073$ MeV, $0.0184$ MeVfm$^{-1}$ and $%
0.1250$ MeV, $0.0352$ MeVfm$^{-1}$, respectively. It could be expected that
transition probabilities at pseudo-crossings 3-4 are much smaller than at
pseudo-crossings 1-2, because the radial velocity decreases at the touching
configuration of the nuclei due to strong frictional effects.

From the neutron level diagram, one can observe that the shells of
individual nuclei dissolve into those of the compound nucleus between $%
7.5-10 $ fm, which agree with the empirical critical distance $r_{cr}=8.01$
fm obtained by Galin's formula \cite{galin}. We will use the value $r_{cr}$
obtained by Galin's formula as the critical distance. It is important to
note that the crossing and the pseudo-crossing of levels occur nearly at the
same relative distances, independent of large variations in the parameter
sets of the TCSM.

\subsection{Nucleus-nucleus potential}

Fig. 2a) shows the total nucleus-nucleus potential for zero orbital angular
momentum of colliding nuclei as a function of the relative distance $r$
between the spherical nuclei (solid curve). We compare the results to those
obtained using a total nucleus-nucleus potential (dashed curve) which is
calculated as the sum of experimental binding energies \cite{Audi} of the
nuclei $(B_{1}+B_{2}-B_{12})$ and a sudden nucleus-nucleus interaction
potential. The $B_{1}$, $B_{2}$ and $B_{12}$ are the experimental binding
energies of the colliding nuclei and the compound nucleus, respectively. The
sudden nucleus-nucleus interaction potential is obtained with the
double-folding method using the Skyrme-type effective density-dependent
nucleon-nucleon interaction and a realistic two-parameter symmetrised
Woods-Saxon function for the density of nuclei \cite{Transp3}. Microscopic
corrections to the double-folding potential are included in the binding
energies.

For large relative distances and for relative distances around the barrier,
the potential calculated with the adiabatic TCSM is similar to the potential
obtained with the double-folding method and small differences are explained
by different nuclear shapes used in the two approaches. The height and the
frequency $\hbar \omega $ of the barrier for both total nucleus-nucleus
potentials are about $26$ MeV and $2$ MeV, respectively. For relative
distances smaller than the position $r_{B}$ of the barrier, the adiabatic
TCSM-potential decreases slower than the double-folding potential because of
a slower decrease of the macroscopic nuclear surface energy (Fig. 2b). Fig.
2a) also shows the adiabatic TCSM-potential around the barrier for the
prolate deformed nuclei in the pole-to-pole configuration (dashed-dotted
curve), where the static deformations (ratio of the semiaxes) of $^{9}$Be 
\cite{Arai} and $^{209}$Bi are taken as $\beta _{1}=1.2$ and $\beta
_{2}=1.05 $, respectively. The deformations also take the effects of
polarisation of the nuclei into account, induced by the mean-field of the
partner-nucleus. In this case, the value of the barrier decreases by $\sim
1.3$ MeV, with approximately the same frequency $\hbar \omega $, and its
position is shifted to $r_{B}\approx 11.5$ fm. The role of this potential
for complete fusion very close to the barrier will be discussed further on.

In Fig.2b), the nuclear surface energy $E_{N}$ is normalised to the nuclear
surface energy $\widetilde{E}_{N}$ for the configuration of the system with
a radius of the neck $R_{neck}=0$. The small increase of this energy at $%
r\approx 12.5$ fm is related to the formation of a neck between the
colliding nuclei. The double-folding potential shows a repulsive core for
small relative distances which reflects the action of the Pauli principle in
the Skyrme-type effective density-dependent nucleon-nucleon interaction. The
numerical values of this sudden potential are similar to the diabatic
potential \cite{Diabatic2} even though both potentials are conceptually and
physically not equivalent. The repulsive character of the diabatic potential
is mainly related to diabatic particle-hole excitations. We would like to
stress that there is practically no difference between the adiabatic and
diabatic potentials of the two nuclei for relative distances larger than the
one corresponding to the touching configuration, because the number of
pseudo-crossings of single-particle levels is very small \cite
{Diabatic2,Diabatic3}.

\subsection{Complete fusion cross section and breakup process}

Fig. 3 shows the complete fusion cross section $\sigma _{fus}$ as a function
of the inverse of the bombarding energy in the center of mass system 1/E$_{%
\text{c.m.}}$. Experimental fusion cross sections (full dots)
for $^{9}$Be + $^{209}$Bi are from \cite{Signorini1}.
It was pointed out in \cite{Signorini1} that the experimental
fusion data for $^{9}$Be + $^{209}$Bi include the complete fusion of $^{8,9}$%
Be. In \cite{Signorini1}, it was assumed that the $^{8}$Be dissociation after
the breakup of $^{9}$Be ($\rightarrow $ n + $^{8}$Be) occurs in a time much
longer than the reaction one. However, we assume the opposite that the decay 
$^{8}$Be$\rightarrow 2\alpha $ occurs automatically and immediately once
$^{9}$Be breaks up before reaching the Coulomb barrier. The
capture either of $2\alpha $ particles by the target $^{210}$Bi 
(in the case of the neutron transfer) or of all projectile fragments 
(n + $2\alpha$) by the target $^{209}$Bi
(in the case of the neutron single-particle motion into the continuum)
can also occur and, therefore, contribute to experimental complete fusion
cross sections. We do not include these events because this is in itself a
complex problem that needs to be investigated further.
The ocurrence of these events as well as the incomplete fusion process
depend on both kinematical aspects of the reaction 
(e.g., the angle between the $2\alpha $ particles, which depends on 
bombarding energy) and dynamical aspects (e.g., motion of $2\alpha $ 
particles in the mean field of the target). Hence, the present calculations
yield a lower limit for complete fusion cross sections.
Comparing this lower limit with the experimental complete fusion
cross sections, we expect that the capture of all individual components of
$^{9}$Be by the target, after the $^{9}$Be breakup, is not very significant.
Our results are also compared with the experimental complete
fusion cross sections (full triangles) obtained for the similar reaction $%
^{9}$Be + $^{208}$Pb \cite{Dasgupta}.

In Fig. 3, we can see that the theoretical complete fusion cross sections
calculated with expression (\ref{fus11}) (dotted curve) agree well with the
experimental complete fusion cross sections. The adiabatic TCSM-potential
for the spherical nuclei is used for these calculations and the results
weakly depend on the value of the critical distance $r_{cr}$. Predictions
(without breakup) of the Glas-Mosel model for this nucleus-nucleus potential
are shown by a solid curve. The theoretical hindrance factor $(1-\overline{%
P_{bup}}($E$_{\text{c.m.}}))$ for complete fusion changes with increasing
bombarding energy from $0.76$ ($24$\% of breakup) for bombarding energies $%
1.05$ times the value of the Coulomb barrier to $0.66$ ($34$\% of breakup)
for bombarding energies about $1.15$ times the value of the Coulomb barrier.
These values agree with the experimental value of $\sim 0.75$ obtained for
the studied reaction \cite{Signorini1} and with the experimental value of $%
0.68\pm 0.07$ reported in \cite{Dasgupta} for the similar reaction $^{9}$Be
+ $^{208}$Pb. The
calculation of complete fusion cross sections with breakup (dotted curve in
Fig. 3) stops at the bombarding energy $1.05$ times the value of the Coulomb
barrier for two reasons, namely the assumption of spherical nuclei is not
suitable for bombarding energies very close to the Coulomb barrier and the
breakup probability (\ref{fus11}) cannot be calculated with the
Landau-Zener formula (\ref{LZ1}) for subcoulomb trajectories because a
radial velocity of the nuclei is not defined in the classically forbidden
region. The arrow in Fig. 3 indicates the value of the Coulomb barrier
of the adiabatic TCSM-potential for the spherical nuclei which is about $40$
MeV. The value of the Coulomb barrier is obtained by subtracting ($%
B_{1}+B_{2}-B_{12}$) from the value of the barrier of the total
nucleus-nucleus potential of Fig. 2a).

In Fig. 3, the dashed-dotted curve shows the complete fusion cross sections
calculated with the Glas-Mosel model (without breakup) for the adiabatic
TCSM-potential with prolate deformed nuclei in the pole-to-pole
configuration. The complete fusion cross section for bombarding energies
very close to the Coulomb barrier can increase due to static deformation and
orientation effects which lower the value of the Coulomb barrier to $\sim
38.7$ MeV. Due to the deformation effects and to the orientation of the
intrinsic symmetry axes of deformed nuclei, the height of the Coulomb
barrier can be larger or smaller than the one obtained with spherical
nuclei. The complete fusion cross section for bombarding energies very close
to the Coulomb barrier can be considered as an averaged value for different
orientations of the intrinsic symmetry axes of the deformed nuclei. It
could be expected that the role of the breakup (Fig. 4, discussed further on),
suppressing the complete fusion (\ref{fus11}), is reduced for
decreasing bombarding energies towards energies below the Coulomb barrier,
and an enhancement of the complete fusion can be a result of large
transmission coefficients for the smaller fusion barriers associated with
certain orientations (e.g., pole-to-pole) of the deformed nuclei.

Fig. 4 shows the breakup probability $P_{bup}$ as a function of the orbital
angular momentum $l$ of captured trajectories $(0\leq l\leq l_{cr})$ for two
values of the bombarding energy E$_{\text{c.m.}}$, namely $42$ MeV (dashed
curve) and $46$ MeV (solid curve). The breakup probability $P_{bup}$ is
calculated as the product of two independent Landau-Zener transitions (\ref
{LZ1}) at the very close pseudo-crossings 1-2, respectively, of Fig. 1. For $%
0\leq l\leq l_{cr}$ and E$_{\text{c.m.}}$= $42$ MeV ($46$ MeV), the breakup
probability $P_{bup}$ decreases by a factor $1.3$ ($1.7$) whereas the
partial capture cross section $\sigma _{cap}$ (\ref{fus2}) increases
by a factor $6.5$ ($10.5$). For a fixed bombarding energy E$_{\text{c.m.}}$,
the breakup probability decreases with increasing orbital angular momentum $%
l $ because the potential energy increases and the radial velocity (kinetic
energy) decreases at the pseudo-crossing points. For a fixed orbital angular
momentum $l$, the radial velocity at the pseudo-crossing points and,
therefore, the breakup probability increases with an increasing bombarding
energy. Results are similar to those obtained when the double-folding
potential determines the trajectories.

\subsection{Elastic and inelastic scattering cross sections}

For a bombarding energy E$_{\text{c.m}.}$ = $46$ MeV, Fig. 5 shows the
scattering function $\theta (l)$ (upper part), the radial distance $r_{\min
} $ of closest approach of the nuclei (middle part), and the dissipated
energy $E_{diss}$ (lower part) as a function of the orbital angular momentum 
$l$ for non-captured trajectories, using two different potentials. For small
orbital angular momenta $l$, the TCSM results (solid curve) are quite
similar to those obtained using the double-folding potential (dot-dashed
curve). Differences mainly appear for large values of $l$ due to the
different tails of the nucleus-nucleus potential. For the adiabatic
TCSM-potential with spherical nuclei, we can observe that inelastic events
are associated with low orbital angular momenta ($12-22\hbar $)
corresponding to angles smaller than the rainbow angle $\theta _{r}$ \cite
{FM1} (where the scattering function $\theta (l)$ has a local maximum). The
dissipated energies for these trajectories are a few MeV. The results do not
depend strongly on the value of the radial friction coefficient $c_{r}^{0}$
(Fig.6, upper part). Since most of these inelastic trajectories show a
turning point inside the radius of the pseudo-crossings 1-2 of Fig. 1, their
classical inelastic cross sections are reduced by the effect of the breakup
of $^{9}$Be. For trajectories with $l=12-16\hbar $, the classical turning
point is relatively far from the position of the pseudo-crossings 1-2 ($%
r\approx 12$ fm) and we use the Landau-Zener approach to calculate the
breakup probabilities $P_{bup}$. In Eq. (\ref{LZ3}), the transition
probabilities $P_{LZ}^{(e)}$ and $P_{LZ}^{(lv)}$ are calculated as in the
complete fusion channel.

Fig. 6 (middle part) shows the angular distribution of inelastic events. The
classical inelastic angular distributions depend weakly on the radial
friction coefficient $c_{r}^{0}$. They slightly change due to small shifts
of the rainbow angle $\theta _{r}$ between $90-95^{\circ }$ corresponding to
dissipated energies between $1.5-3$ MeV for the scattered trajectories with
the rainbow angle $\theta _{r}$. Classical inelastic cross sections are very
large in the neighbourhood of the rainbow angle (rainbow scattering) and the
cross section becomes infinite at $\theta _{r}$. Since the projectile $^{9}$%
Be has no excited bound states, the inelastic events are related to the
excitation of the target $^{209}$Bi. The breakup of $^{9}$Be reduces the
classical inelastic cross sections by $29-34$\% for scattering angles
smaller than the rainbow angle $\theta _{r}$, but their values still remain
appreciable. The experimental point is extracted from \cite{Signorini2} and
corresponds to the maximum of the angular distribution of the inelastic
collective (vibrational) multiplet $3^{-}$ of $^{209}$Bi at around $2.6$ MeV
for a scattering angle of $95^{\circ }$. This multiplet could not be
observed at angles smaller than $\sim 90^{\circ }$, which was explained by
an increasing background $\emph{most probably originating from the tail of
the elastic peak}$. The large values of the elastic cross section (one-two
orders of magnitude larger than the inelastic one) for angles smaller than $%
\sim 90^{\circ }$ support this claim (Fig. 6, lower part). This reason could
also explain that all the other $^{209}$Bi inelastic channels were observed
with negligible cross sections \cite{Signorini2}.

We have assumed that a corresponding classical trajectory is associated with
the peak of the experimental angular distribution for this inelastic
channel. Moreover, the classical trajectory corresponds to the mean
trajectory. In fact, the width of the experimental angular distribution for
a given inelastic channel is caused by quantal effects and statistical
fluctuations \cite{Fobrich1}. Although the experimental point of Fig. 6
(middle part) cannot be reproduced in the present classical treatment, we
would like to note that the classical trajectory corresponding to the
dissipated energy of $2.6$ MeV has a scattering angle very close to the
experimental one being in the vicinity of the rainbow angle $\theta _{r}$.
In reality the classical singularity at the rainbow angle is removed due to
quantal effects, statistical fluctuations and also experimental broadening 
\cite{Fobrich1}. For example, the inelastic angular distribution in the
semiclassical approximation for rainbow scattering \cite{FM1} using the Airy
formula shows a smooth exponential decrease from $\sim 200$ mb to $\sim $ $%
100$ mb with increasing scattering angles around the rainbow angle $\theta
_{r}$. However, the semiclassical method of scattering \cite{FM1} is not
appropriate for the reaction studied here because the reduced mass $\mu $ of
the system is very small and the value of de Broglie wavelength $h/\sqrt{%
2\mu E}$ of nuclei in the relative motion is comparable to their nuclear
radii. The experimental value in Fig. 6 (middle part) could be on the ``dark
side'' ($\theta >\theta _{r}$) of the angular pattern of rainbow scattering.
This experimental value of the inelastic cross section can, however, well be
reproduced by the coupled channels method \cite{Signorini2}, where the
relative motion of the nuclei is treated quantum mechanically.

Fig. 6 (lower part) shows the angular distribution of elastic events,
normalised to the Rutherford cross section $\sigma _{R}$. For smaller
scattering angles, the elastic cross section obtained with the
adiabatic-TCSM potential slightly deviates from the Rutherford cross section
due to the deviation of the tail of the adiabatic TCSM-potential from the
Coulomb tail for large $r$. The increasing values around $\sim 90^{\circ }$
occur because the scattering angles are near the rainbow angle $\theta _{r}$%
. The effect of the breakup in the elastic scattering is not considered
because the turning point for elastic trajectories is outside the radius of
the pseudo-crossings 1-2 of Fig. 1. Results obtained when the double-folding
potential determines the elastic trajectories are closer to the experimental
data.

For decreasing bombarding energies, the reduction of the projectile
penetration leads to a decrease of the dissipated energy for inelastic
events. We found that the rainbow angle $\theta _{r}$ in the scattering
function $\theta (l)$ disappears for E$_{\text{c.m.}} \approx 44$ MeV. The
scattering angle for inelastic trajectories (E$_{diss} \geq 0.5$ MeV)
becomes larger than $95^{\circ }$ increasing to larger angles with
decreasing bombarding energies. Inelastic cross sections become smaller with
decreasing bombarding energies because the increase of $d\theta (l)/dl$ is
larger than the increase of the prefactor of (\ref{scat2}) proportional to E$%
_{\text{c.m.}}^{-1}$. Moreover, the breakup probability $P_{bup}$ of $^{9}$%
Be decreases with decreasing bombarding energies for inelastic trajectories
(E$_{diss} \geq 0.5$ MeV) where the radial distance $r_{\min }$ of closest
approach is smaller than $\sim 12$ fm (inelastic events with smaller values
of $l$). This probability changes with decreasing bombarding energy from $%
0.34- $ $0.29$ ($l=12-16\hbar $) for bombarding energies $1.15$ times the
value of the Coulomb barrier to $0.2-0.11$ ($l=8-10\hbar $) for bombarding
energies about $1.05$ times the value of the Coulomb barrier. As in the
complete fusion channel, the effect of the breakup of $^{9}$Be on inelastic
cross sections is diminished with decreasing bombarding energies.

\section{Summary and conclusions}

The breakup of the $^{9}$Be projectile on the $^{209}$Bi target at
bombarding energies above and near the Coulomb barrier has been studied in
the adiabatic two-center shell model approach as well as its effect on the
complete fusion and scattering processes. The neutron level diagram of the
two-center shell model reveals two very close pseudo-crossings between the
state $j_{z}=3/2$ of the last neutron of $^{9}$Be and two unoccupied states
of $^{209}$Bi with the same projection of the single-particle total angular
momentum $j_{z}$, shortly before the colliding nuclei reach the Coulomb
barrier. A radial velocity-dependent transition of the last neutron of $^{9}$%
Be into the continuum and to $^{209}$Bi at these pseudo-crossings of
molecular levels leads to the most favourable breakup channel of $^{9}$Be ($%
^{9}$Be $\rightarrow$ n + $^{8}$Be). Since the nucleus $^{8}$Be has no bound
states, the Coulomb field of the target ($^{209}$Bi or $^{210}$Bi) in its
vicinity is very strong and the relative motion of the nuclei is considered
adiabatic, we assumed that the decay $^{8}$Be $\rightarrow 2\alpha $ occurs
automatically and immediately. Breakup probabilities depending on the
bombarding energy and the orbital angular momentum have been calculated with
the Landau-Zener approach and trajectories obtained from a classical model
using the adiabatic TCSM-potential along with phenomenological friction
coefficients.

The effect of this breakup channel on complete fusion and scattering
(elastic and inelastic) cross sections was studied. From the value obtained
with the Glas-Mosel model without breakup, the complete fusion cross section
reduces by $34\%$ for the bombarding energy $1.15$ times the value of the
Coulomb barrier and by $24\%$ for the bombarding energy $1.05$ times the
value of the Coulomb barrier. Calculated complete fusion cross sections
agree with the experimental data. The breakup probability of $^{9}$Be
decreases with decreasing bombarding energies towards the Coulomb barrier,
with a hindrance factor of $0.66-0.76$ for complete fusion which is similar
to the experimental one of $\sim 0.75$ obtained for the studied reaction and
to the experimental value of $0.68\pm 0.07$ for the similar reaction $^{9}$%
Be + $^{208}$Pb. For a fixed bombarding energy, the breakup probability
increases with decreasing orbital angular momentum and it becomes maximal
for a central collision.

The effect of the breakup of $^{9}$Be on inelastic channel is similar to its
effect on complete fusion channel. Inelastic cross sections decrease with
decreasing bombarding energies. Results obtained for E$_{\text{c.m}}$ $=46$
MeV qualitatively explain the experiment. The results are similar to those
obtained when the double-folding potential determines the trajectories.

In general, the results indicate that absorption channels for bombarding
energies very close to the Coulomb barrier observed in experiment \cite
{Signorini2} can also be associated with orientation and static deformation
effects of $^{9}$Be which lower the value of the Coulomb barrier and
increase the fusion cross section. Our results also support the molecular
description of the highly asymmetric reaction $^{9}$Be + $^{209}$Bi in the
framework of the two-center shell model.
It would be interesting to study other reactions with $^{9}$Be but a more
general two-center shell model \cite{Nuhn} would be needed in order to describe, e.g.,
arbitrary orientations of the intrinsic symmetry axes of deformed nuclei. 
Calculations for the studied reaction using a three-body model within a 
coupled discretised continuum channels (CDCC) formalism are in progress.

{\bf Acknowledgments}

The authors would like to thank Dr. G.G. Adamian and Dr. N.V. Antonenko for
a careful reading of the manuscript, fruitful discussions and suggestions,
and Mrs. N. Diaz-Torres for her help in preparing the paper. We would like
particularly to thank Prof. C. Signorini and Dr. M. Dasgupta for their
experimental data. UK support from the EPSRC grant GR/M/82141 is acknowledged.

\pagebreak

\newpage
\begin{figure}
\begin{center}
\epsfig{file=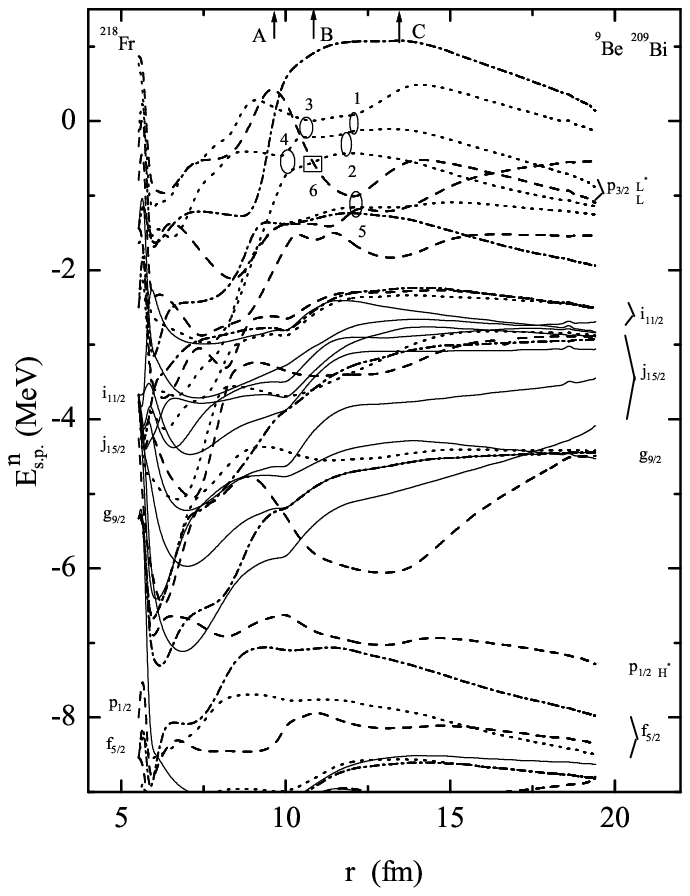,width=0.75\textwidth}
\end{center}
\caption{Neutron levels for $^{9}$Be + $^{209}$Bi $\rightarrow $ $^{218}$Fr
as a function of the separation $r$ between the nuclei. Levels are
characterised by the total angular momentum projection $j_{z}$ on the
internuclear axis: $j_{z}=1/2$ (dashed curves), $j_{z}=3/2$ (dotted curves), 
$j_{z}=5/2$ (dashed-dotted curves) and other values (solid curves). See text
for further details.}
\end{figure}

\newpage
\begin{figure}
\begin{center}
\epsfig{file=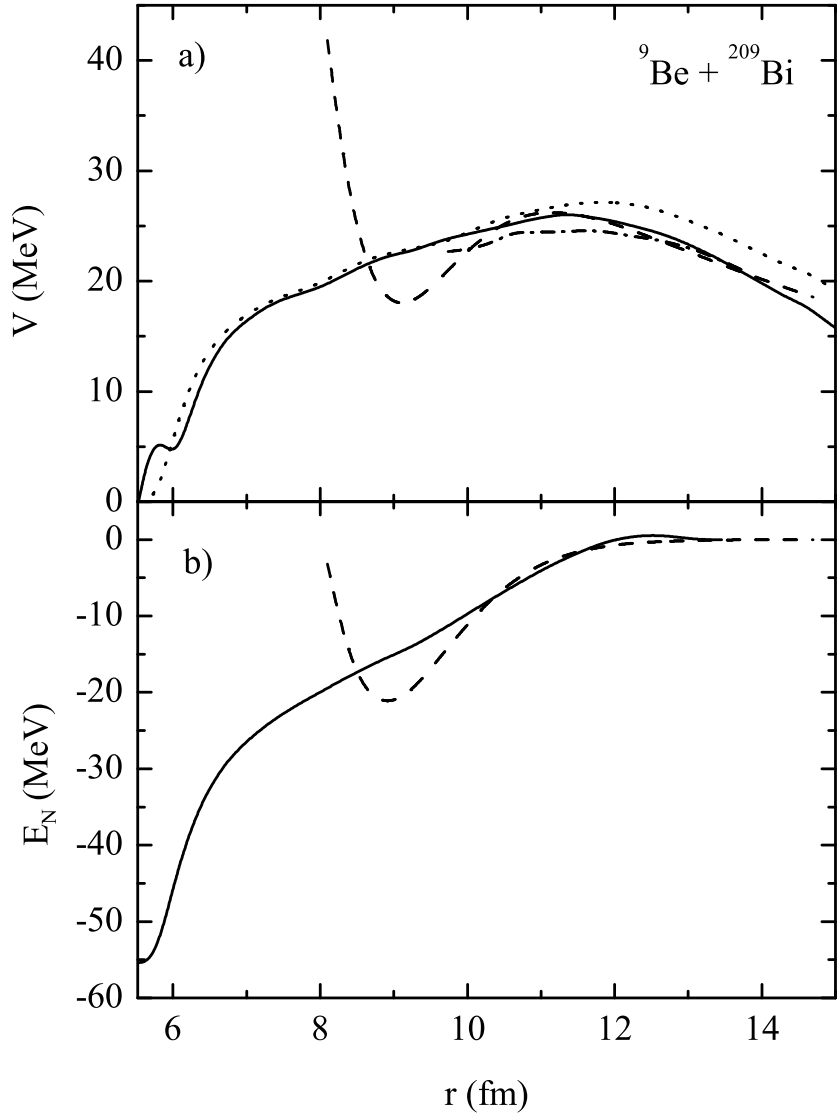,width=0.75\textwidth}
\end{center}
\caption{a) The angular momentum independent part of the total
nucleus-nucleus potential as a function of the separation $r$ between the
nuclei. The adiabatic TCSM-potential with spherical nuclei and the sudden
double-folding potential are shown by solid and dashed curves, respectively.
The smooth macroscopic part of the adiabatic TCSM-potential is presented by
a dotted curve. The adiabatic TCSM-potential for the prolate nuclei in the
pole-to-pole configuration is shown by a dashed-dotted curve. b) The nuclear
surface energy $E_{N}$ with spherical nuclei is shown by a solid curve. The
nuclear double-folding potential is presented by a dashed curve. See text
for further details.}
\end{figure}

\newpage
\begin{figure}
\begin{center}
\epsfig{file=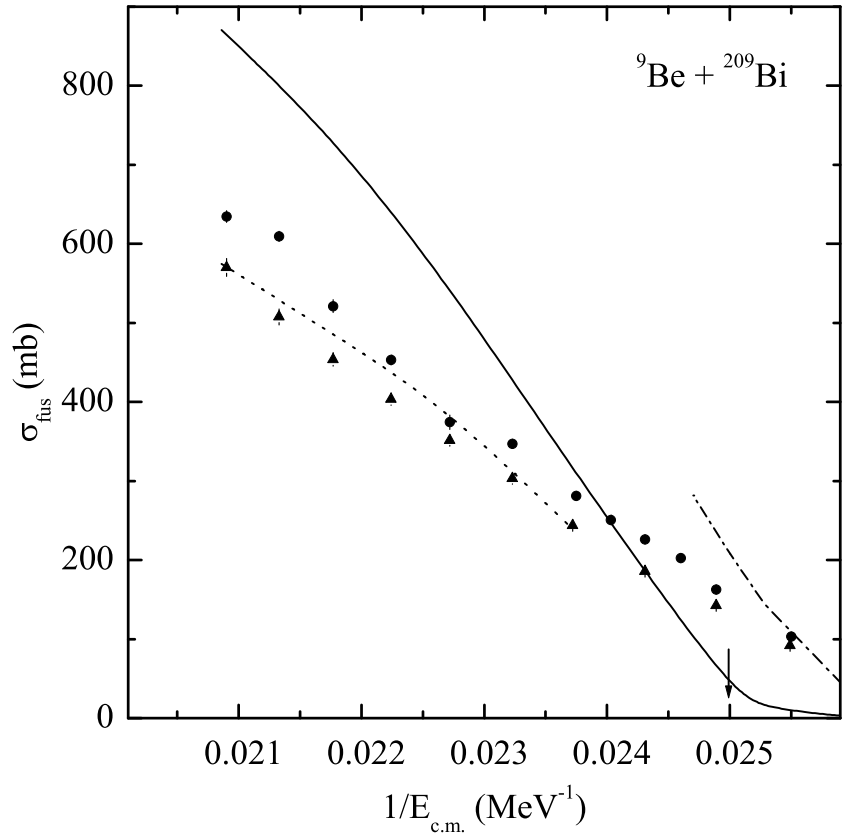,width=0.75\textwidth}
\end{center}
\caption{Complete fusion cross sections $\sigma _{fus}$ as a function of the
inverse of the bombarding energy 1/E$_{c.m.}$ in the center of mass system.
Predictions (without breakup) of the Glas-Mosel model for the adiabatic
TCSM-potential with spherical nuclei (see Fig. 2a) are shown by a solid
curve and values taken into account the breakup of $^{9}$Be, by a dotted
curve. Experimental data for the studied reaction (full dots) are 
from \protect\cite{Signorini1} and for $^{9}$Be + $^{208}$Pb (full triangles) 
from \protect\cite{Dasgupta}. Complete fusion cross 
sections $\sigma _{fus}$ calculated with the Glas-Mosel model 
(without breakup) for the adiabatic TCSM-potential with prolate nuclei 
(see Fig. 2a) are presented by a dashed-dotted curve. 
See text for further details.}
\end{figure}

\newpage
\begin{figure}
\begin{center}
\epsfig{file=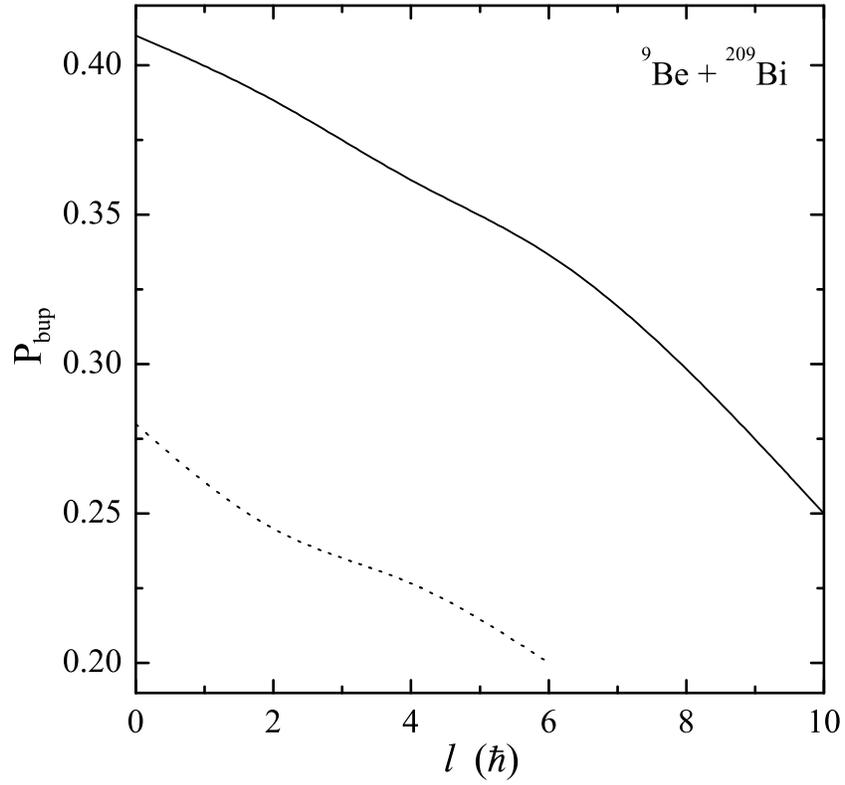,width=0.75\textwidth}
\end{center}
\caption{Breakup probabilities $P_{bup}$ as a function of the orbital angular
momentum $l\hbar $ ($0\leq l\leq l_{cr}$) for captured trajectories obtained
for the adiabatic TCSM-potential with spherical nuclei (see Fig. 2a). Values
for E$_{c.m.}=42$ MeV and E$_{c.m.}=46 $ MeV are shown by dashed and solid
curves, respectively. See text for further details.}
\end{figure}

\newpage
\begin{figure}
\begin{center}
\epsfig{file=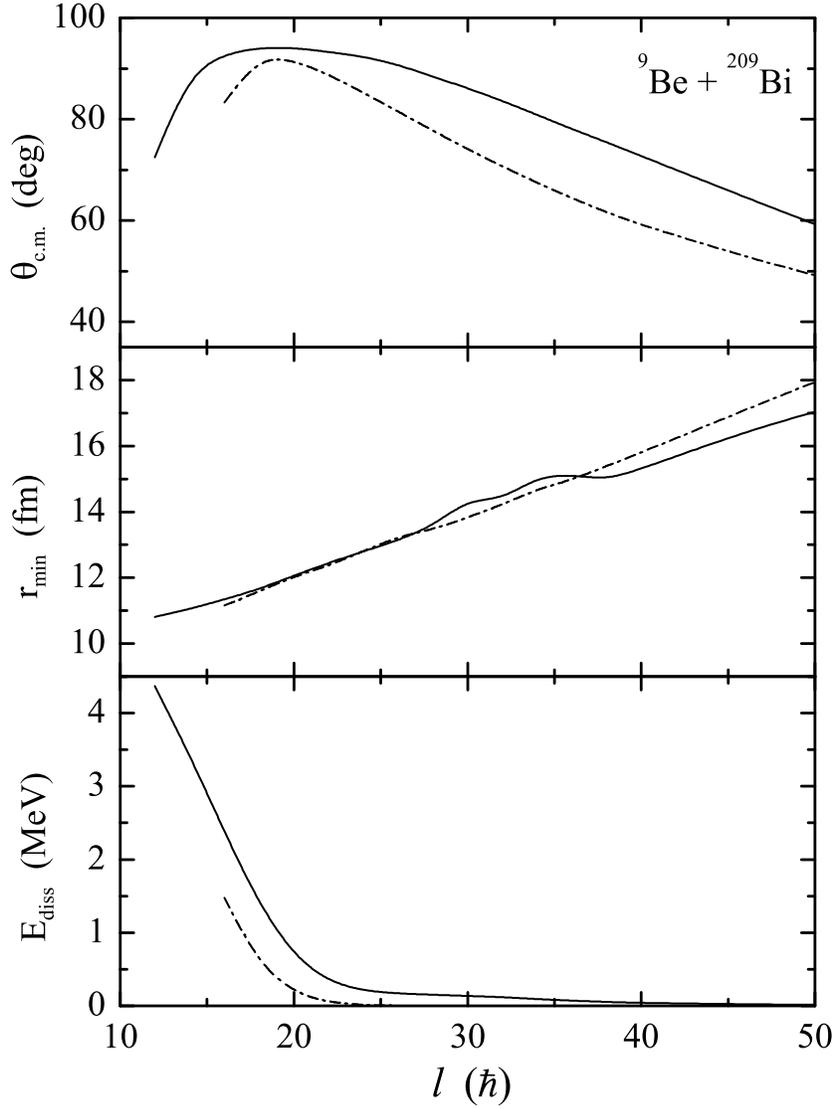,width=0.75\textwidth}
\end{center}
\caption{Scattering function $\theta _{c.m.}(l)$ (upper part), distance of
closest approach $r_{\min }$ of the nuclei (middle part) and dissipated
energy $E_{diss}$ (lower part) as a function of the orbital angular momentum 
$l\hbar $ ($l>l_{cr}$) for non-captured trajectories at $E_{c.m.}=46$ MeV.
Values obtained for the adiabatic TCSM-potential with spherical nuclei (see
Fig. 2a) and for the double-folding potential are shown by solid and
dot-dashed curves, respectively. See text for further details.}
\end{figure}

\newpage
\begin{figure}
\begin{center}
\epsfig{file=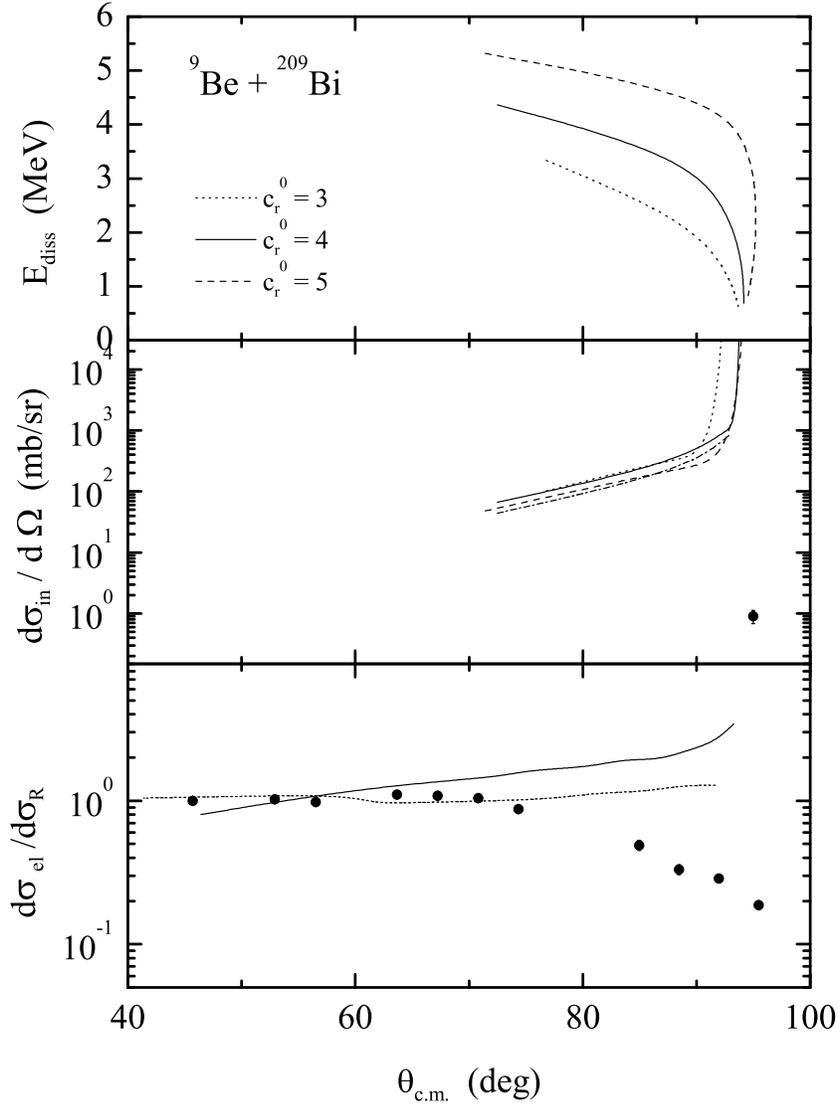,width=0.75\textwidth}
\end{center}
\caption{At $E_{c.m.}=46$ MeV, dissipated energies $E_{diss}$ (upper part)
and inelastic classical cross sections $d\sigma _{in}/d\Omega $ (middle
part) as a function of the scattering angle $\theta _{c.m.}$ for inelastic
non-captured trajectories obtained with the adiabatic TCSM-potential for
spherical nuclei (see Fig. 2a) by using different values of the radial
friction coefficient $c_{r}^{0}$ in units of $10^{-23}$ MeV$^{-1}$s. For a
radial friction coefficient $c_{r}^{0}=4\times 10^{-23}$ MeV$^{-1}$, the
inelastic angular distribution taking into account the breakup of $^{9}$Be
is shown by a dashed-dotted curve. The experimental point (full dot) is from 
\protect\cite{Signorini2}. 
Elastic angular distributions $d\sigma _{el}/d\sigma _{R}$ (lower part), 
normalised to the Rutherford cross section $\sigma _{R}$, obtained for 
the adiabatic TCSM-potential with spherical nuclei and for the double-folding 
potential are shown by solid and short-dashed curves,
respectively. Experimental data (full dots) are 
from \protect\cite{Signorini2}. See text for further details.}
\end{figure}

\end{document}